\documentclass[aip,graphicx,reprint,floatfix]{revtex4-1}
\usepackage{amsmath,amssymb,graphicx,color,wasysym}
\usepackage{textcomp,ulem}
\usepackage{hyperref}

\draft

\begin{document}

\title{Rheology and structure of a suspension of deformable particles in plane Poiseuille flow}


\author{Vassanti Audemar}
\address{Universit\'e Grenoble Alpes, CNRS, LIPhy, F-38000 Grenoble, France}
\address{Universit\'e Paris Saclay, I2BC, Paris, France}

\author{Thomas Podgorski}

\address{Universit\'e Grenoble Alpes, CNRS, LIPhy, F-38000 Grenoble, France}
\address{Universit\'e Grenoble Alpes, CNRS, Grenoble INP, LRP, F-38000 Grenoble, France}

\author{Gwennou Coupier}\email{gwennou.coupier@univ-grenoble-alpes.fr}

\address{Universit\'e Grenoble Alpes, CNRS, LIPhy, F-38000 Grenoble, France} 

 \email{gwennou.coupier@univ-grenoble-alpes.fr}
\date{\today}

\begin{abstract}
We present an experimental study of the rheology and structure of a confined suspension of deformable particles flowing in a quasi-two-dimensional Poiseuille flow. Thanks to a precise microfluidic viscosimetry technique combined with measurements of concentration profiles, our study provides the first experimental confirmation {with three-dimensional particles} of a strong relationship between structuring effects and rheology, previously only reported in numerical simulations of purely two-dimensional systems. In conditions where strong structuring effects take place due to confinement, the evolution of the effective viscosity with particle concentration (here, red blood cells) shows a remarkable succession of ranges of rapid growth and plateaus that are associated to qualitative transitions in the structure of the suspension.

\end{abstract}

\maketitle

\section{Introduction}
The rheological properties of complex materials may differ drastically depending on the scale at which they are probed. When a specific length scale is imposed by the flow geometry, such as when confining the material, the bulk properties of the material are often insufficient to properly describe the flow behaviour when this length scale becomes comparable to the characteristic length of the microstructure. A well-known example is the F\r{a}hr{\ae}us-Lindqvist effect in blood flows, where the apparent viscosity in tube flows decreases with the tube diameter due to structuring of the suspension and the formation of a cell-free layer near walls \cite{Lindqvist31}.

Even for simple molecular fluids with a Newtonian behaviour at the continuum scale, the molecular scale sets a threshold below which the fluid cannot be considered any more as a continuous material: layering transitions marked by jumps in the normal force have been observed\cite{horn81}, or dependency of the effective shear viscosity with the gap size \cite{bureau10,Maali06}.

Three orders of magnitude above, suspensions of micron-size particles have more recently been shown to exhibit such connections between layering effects and rheology. The actual picture is still sparse ;  while there seems to be no clear correlation between rheology and structure for colloidal suspensions \cite{xu13}, non-colloidal suspensions of hard spheres exhibit jumps in shear viscosity associated with varying the gap of the Couette geometry: simulations \cite{aerov15,fornari16} show that for gaps of order a few particle diameters, the effective viscosity oscillates with the gap size, with a period equal to particle diameter, and viscosity minima are found for gaps that are multiple of this particle diameter. In a less confined situation \cite{yeo10}, a  decrease in viscosity is also associated with ordering transitions, when they come to occur. {Other studies have highlighted some ordering in the plane of shear, with no clear signature on the dissipation \cite{blanc2013,deboeuf2018}.}

The above mentioned systems of rigid particles exhibit layering transitions when the particle concentration is such that steric effects trigger ordering. Deformable particles influence each other over larger distances through hydrodynamic interactions \cite{loewenberg97,singh09,lac08,pranay10,gires14}, thus opening the possibility for suspension structuring at relatively lower concentrations.


Two-dimensional (2D) numerical simulations of a suspension of vesicles (drops encapsulated by an incompressible membrane) have highlighted such structuring effects under imposed simple shear flow. Plateauing of the effective  viscosity is observed periodically as a function of cell concentration at fixed gap size \cite{thiebaud14,shen17,Naitouhra2019}. This feature coincides with the self-organization of the particles into discrete layers separated by layers of suspending fluid, until a new layer is created at a critical value of the concentration.  2D simulations of capsules in Poiseuille flow have recently exhibited similar phenomenon: a marked increase in viscosity is associated to the transition from single file to double file\cite{feng21}. {Another study on a similar system has also highlighted, yet on a sparse set of data, the influence of cell organization on effective rheology \cite{lazaro19}}.Since the 2D nature of these configurations leads to strong topological constraints and  interactions that favor the alignment of particles and their organization into regular layers, one can ask to what extent these structures and their impact on rheology are significant in three-dimensional (3D) suspensions, simulated or experimental. Due to the relatively high computational cost of 3D simulations, systematic studies of the structure-rheology relationship in 3D channel flows of suspensions are still scarce \cite{fedosov14b}. Experimentally, the simultaneous measurement of subtle changes in the structural organization of suspensions in channels and of related viscosity variations in ranges of particle volume fractions between 0 and 50 \% is a considerable challenge.

In this paper, we contribute to this important structure-rheology link in confined suspensions, which has important consequences on flow and particle transport in confined channels and networks, by exploring experimentally the rheological behaviour of a confined model suspension of deformable particles. To that end, we consider red blood cells, which are conveniently monodisperse enough to be a model system and whose mechanical properties and dynamics have received a lot of attention in recent experimental, theoretical and computational works. Red blood cells exhibit collective dynamics that can lead to the expected layering behaviour in confined flow, a phenomenon that we control by varying the viscosity of the external fluid. Through measurements of the effective viscosity of the suspension and of concentration profiles and by varying RBC volume fraction, we show correlations between viscosity variations and layering of the suspension when the viscosity of the suspending fluid favors the formation of coherent structures through hydrodynamic interactions, that are reminiscent of observations made in 2D simulations in Poiseuille flow \cite{feng21}.

 \section{Experimental set-up}

   \begin{figure*}[t]
	\centering 		
\includegraphics[width=\textwidth]{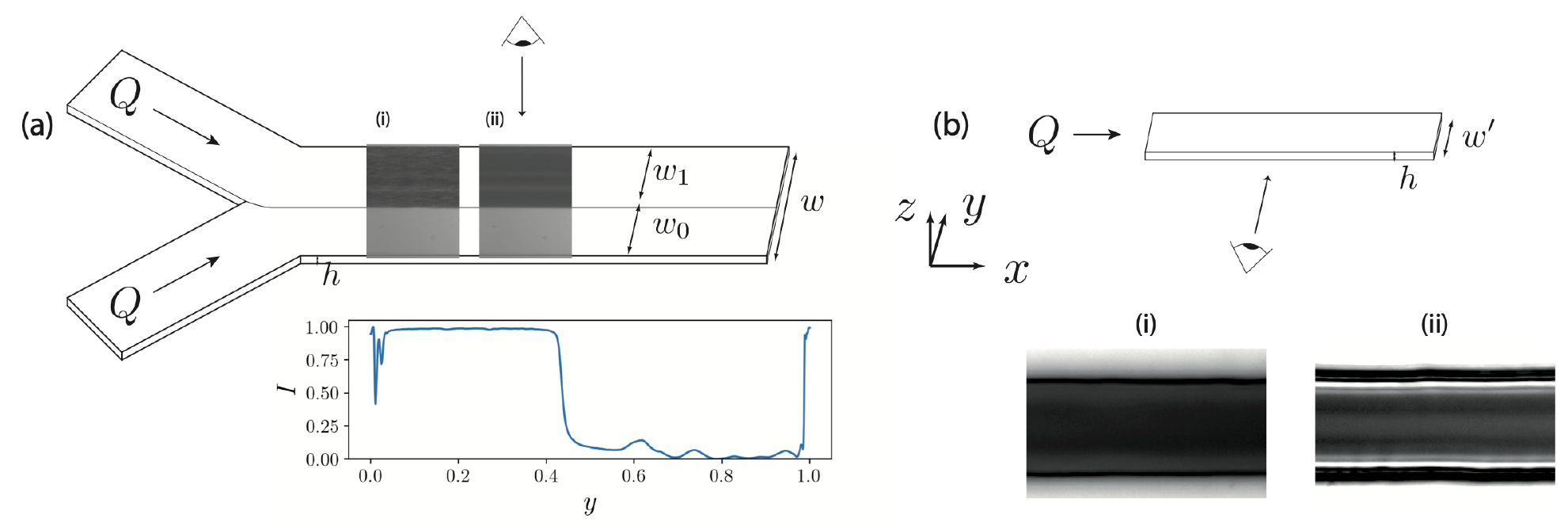}
	\caption{(a) Principle of the microfluidic viscosimeter. The channel is very flat, with $w=50 h$. A fluid of known viscosity $\eta_0$ spans over a width $w_0$ while the other fluid (here, the suspension of red blood cells) spans over a width $w_1$. An example of {(i) a raw image and (ii) a time-averaged image}, viewed along the $z$ axis, are shown inside the channel. The intensity profile along the $y$ axis is calculated by averaging also in the flow direction $x$. Inset shows such a profile, which has been normalized by its extreme values. (b) A different set-up allows for visualisation of the associated structure in the shear direction. The width of the channel is $w'\ll w$, but a flat configuration is preserved: $w'=3.25 \times h$. Examples of time-averaged pictures are shown below for (i) $\eta_0=1.6$ mPa.s and $\Phi_T=0.204$, (ii) $\eta_0=5.8$ mPa.s and $\Phi_T=0.137$ {(contrast enhanced for illustrative purposes)}. \label{fig:schema}}
\end{figure*}
 
Red blood cell suspensions are prepared following standard techniques for in-vitro studies. Blood was supplied by the Etablissement Fran\c{c}ais du
Sang (EFS). It was taken from healthy volunteers {and stored in tubes containing citrate to prevent coagulation}. Blood
samples were first washed 3 times in a solution of
phosphate-buffered saline solution (PBS : 9 mM Na2HPO4, 1.3 mM NaH2PO4, 140 mM NaCl, pH 7.4) and red blood cells were separated from other components of blood ({platelets, white cells, plasma, citrate}) by centrifugation.  From the pellet, successive dilutions were performed using 3 suspending solutions with different viscosities. Following an initial suggestion by Roman \textit{et al.}\cite{roman12}, which has been fruitfully exploited in several subsequent studies \cite{roman16,Shen16,losserand19,minetti19,merlo22}, the density of the suspending solutions was adjusted with Optiprep (iodixanol solution, Axis Shield) to match the average density of red blood cells and thus avoid sedimentation during the experiments. Three solutions where prepared, respectively with  (a) water  68.5 \%, Optiprep 31.5 \%, PBS, BSA 0.1 \%, of viscosity 1.6 mPa.s at 20$^\circ$C ; (b) water  72.5 \%, Optiprep 27.5 \%, Dextran 100 kDa 6.5 \%, PBS, BSA 0.1 \%, of viscosity 5.8 mPa.s at 20$^\circ$C ; (c) water  75 \%, Optiprep 25 \%, Dextran 100 kDa 10 \%, PBS, BSA 0.1 \%, of viscosity 8.1 mPa.s at 20$^\circ$C. BSA (Bovine Serum Albumin)  aims at maintaining an initial coating of the walls of the microfluidic system to prevent cell adhesion. The concentration of Dextran 100 kDa (from \textit{Leuconostoc mesenteroides}, Roth) is such that it does not induce RBC agregation\cite{neu2002}.

Bulk shear viscosity measurements were performed in a cone-plate rheometer (Anton Paar MCR  301) at fixed shear rate 1000 s$^{-1}$, in a water-saturated atmosphere to prevent evaporation.  The cone-plate geometry was cleaned between each sample.

Shear rheology of highly confined liquids is a difficult task as it requires to maintain a constant gap thickness while measuring potentially low stresses. An attempt to use an elaborate small-gap rheometer that had proven useful in other systems \cite{baik11,jofore15} showed that low viscosities of our suspensions and their small variations reach the sensitivity limit of the instrument. Visualisation of the structure of the suspension is also problematic in such a shear flow chamber where side view cannot be implemented. Using confocal microscopy to scan the sample in planes parallel to the walls is tricky as soon as the RBC volume fraction reaches a few percent, due to the high refraction of light by red blood cells.

We turned to a simple yet powerful microfluidic design proposed by  Galambos and Forster \cite{Galambos1998ANOM}, and developed for the study of different systems \cite{guillot06,solomon14,solomon16}. It allows viscosity measurements of a fluid under a quasi-2D Poiseuille flow using co-flowing laminar streams. Such a set-up has already been used to measure the viscosity of low stress fluids such as an active suspension of \textit{E. coli} bacteria  \cite{gachelin13}, or blood \cite{Kang16,KIM18}. In these latter studies, the confinement of the suspension was not as important as here (50 $\mu$m Vs 20 $\mu$m).  More precisely, we consider a Y-shaped Hele-Shaw cell, built on demand in glass by Klearia. Each inlet is fed by two different fluids, a Newtonian one of known viscosity $\eta_{0}$ and another one whose viscosity $\eta_{1}$ is to be measured (see Fig. \ref{fig:schema}(a)).   Both flows are driven at an identical flow rate $Q=10\,\mu$L/min, corresponding to a maximum wall shear rate of 5000 s$^{-1}$, thanks to a two-channel syringe pump (KDS Legato 180). The two flows meet in a very flat channel of height $h=20\,\mu$m , width $w=1$ mm and length $L=15$ mm, such that shear is mainly in the direction $z$ of the thickness of the channel. The choice of glass as material is dictated by the necessity to avoid uncontrolled swelling of the channels under pressure\cite{gervais06}, which would be an issue for both the deduction of the viscosity of the unknown sample from the image analysis, and the analysis in term of suspension structuring.

Thanks to the low Reynolds number ({of order 0.1}), the position of the interface between the two fluids reaches a steady position along the flow axis $x$ at a distance of order $h$, after which the width $w_i$, $(i=0,1)$ occupied by each fluid is measured.
When the interface between the two fluids is far from the lateral walls, solving the Stokes equation with the no slip condition at the channel walls, the velocity continuity between the two fluids streams and the low aspect ratio, leads to a simple relationship between viscosities and the widths of the two parallel streams\cite{Galambos1998ANOM}: $ \eta_1/\eta_0 =w_{1}/w_{0}$. The viscosity of the tested fluid is therefore deduced from the measurement of the position of the interface. The precision in the measurement is associated with the precision in the determination of the positions of the {lateral} walls and of the interface between the two liquids. Here, we measure this position before shear-induced-diffusion blurs the interface \cite{grandchamp13}. {Observations are made around 5 mm after the two flows merge. Since the cells have already flown in the inlet branch of length at least 5 mm, this is highly enough to ensure the cell structure is steady:  two different studies \cite{zhou2020,losserand20} show that for a confinement of 20 microns, 2 mm are sufficient to establish a steady structure.}  For each sample, a time of 10 minutes was respected between the beginning of the injection and the time of measurement to allow stabilization of the flow. This relaxation time is a well-known direct consequence of driving a flow by imposing its rate through channels with a high hydraulic resistance \cite{Lake2017} and  has been reduced as much as possible by an adequate choice of rigid materials such as small glass syringes (Hamilton  1002TLL 2.5 mL) and PTFE tubing. The chip was placed on an inverted microscope (Olympus IX 71) for imaging.

Before each experiment, the channel was filled with distilled water and placed under vacuum to remove any trapped bubble that would strongly increase equilibration times. The channel was then filled with a PBS solution with 1\% of BSA and kept in a refigerator overnight. This allowed albumin to coat the surfaces in such a way that no cell adhesion would occur during experiments by preventing contact between RBCs and glass. Syringes were also treated. Then, the channel was filled with the reference fluid of viscosity $\eta_0$, which is the
fluid (a), (b) or (c) that was used to suspend cells
and a reference image was taken. The suspension to be studied was then injected through the other inlet and both fluids were driven at the same flow rate. 
The channel was rinsed with the same reference fluid when switching between different samples prepared with the same considered medium but a different red blood cell concentration. When switching to a different suspending medium, the channel was completely cleaned and refilled.

For each sample, a short movie was recorded once the interface reaches a steady position, and the time-averaged  image of the stack was calculated using ImageJ in order to reduce noise and small fluctuations. The background (reference) image was then subtracted. To detect the position of the interface between the test fluid and the reference fluid, the intensity profile across the channel ($y$ direction) was averaged over the dimension of the image along the flow direction $x$ (see Fig. \ref{fig:schema}(a)),  and fitted with an error function with IDL software. The position of the interface was chosen at 85\% of the height separating the two plateaus of the error function, yielding the width $w_0$ and $w_1$ of each stream. The viscosity ratio $ \eta_1/\eta_0 =w_{1}/w_{0}$  was then deduced.  {The inaccuracy in the detection of the interface and side walls is estimated to about $\pm$2 pixels, which leads to a maximum error on $w_1/w_0$ of about $\pm 1\%$,
 a value that is used for error bars on the graphs.}\\

For each sample, its volume fraction, called reservoir volume fraction $\phi_R$, was measured by centrifugation of the samples in glass haematocrit capillary tubes. The tube volume fraction $\phi_T$ is the local volume fraction of the suspension in the flow. In confined flows, it generally differs from the reservoir volume fraction, a phenomenon known as F\r{a}hr\ae{}us effect for blood\cite{pries92}. This is a consequence of the non homogeneous distribution of cells across the channel thickness and the formation of a cell-free layer near walls, leading to an average velocity of red blood cells higher than the average fluid velocity.
Tube volume fraction was determined by an effective absorbance measurement of flowing red blood cells (Beer-Lambert law), a method that has proven to be reliable up to rather high volume fractions levels \cite{pries83,roman16,mantegazza20,merlo22}. The effective absorption coefficient was calibrated on diluted samples of cells, that could be counted individually. The conversion into cell volume fraction is then done by considering 90 $\mu$m$^3$ as the average red blood cell volume\cite{McLaren87}.

\begin{figure}
	\centering 		
\includegraphics[width=0.5\textwidth]{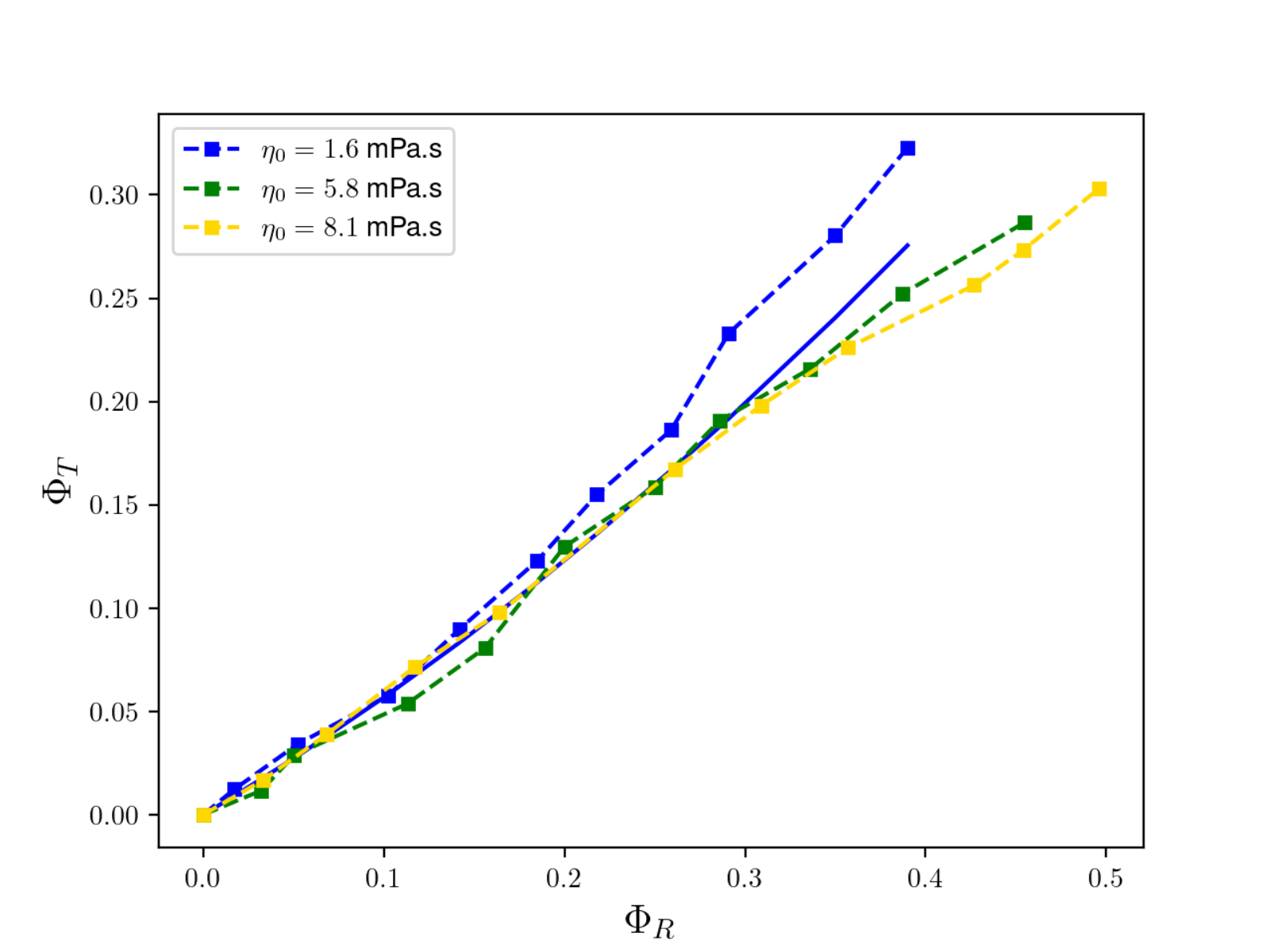}
	\caption{Tube volume fraction $\Phi_T$  in the micro-viscosimeter as a function of reservoir volume fraction $\Phi_R$, for three different initial viscosities of the suspending fluid.  Full line corresponds to the relationship given by Pries {\it et al.} \cite{pries92}, for blood in cylindrical channels of 20 $\mu$m, characterizing the F\r{a}hr{\ae}us effect \cite{Fahraeus1929}.\label{fig:PhiT}}
\end{figure}

The micro-viscosimeter does not allow for  an easy evaluation of the structure of the suspension in the shear direction $z$.  Direct determination requires to observe the flow in the $y$ direction, which is not possible due to the very high aspect ratio of 20 $\mu$m $\times$ 1~mm of this channel. As a compromise, we injected the same suspensions of cell  in straight channels made in a PDMS chip of rectangular cross section  with the same thickness $h=20\, \mu$m  in the $z$ direction and a width $w'=65 \mu$m in the $y$ direction. Its  orientation in the microscope is such that the optical axis is now the $y$ axis (see Fig. \ref{fig:schema}(b)). {Observation is made at a distance of at least 5 mm from the inlet, to ensure a steady profile is observed}. With a PDMS technology, a higher aspect ratio $w'/h$ would imply non straight walls upon bonding of the PDMS matrix on the sealing plate, and in addition a too high aspect ratio would lead to optical aberrations due to reflections and refraction of incident light by the walls, whatever their quality and material. 
In our case, while we are not exactly in a situation where $w'\gg h$, we still have $w'/h=3.25$, meaning that the flow and the structure of the suspension are dominated by the smallest dimension $h$ that sets the main direction of velocity gradients. The particle distribution and layering effects are therefore expected to be very similar{, though probably smoother,} to the flow structure taking place in the microfluidic viscometer. { Note that while structuration in the vorticity direction has been reported for highly confined plane Poiseuille flow of RBC suspensions \cite{iss2019}, it does not seem to take place in our case where the thickness is twice as large: no structure is visible from the direct vizualisation in the viscosimeter (Fig. \ref{fig:schema}(a)).}
The relationship between the volume fraction $\Phi_T$ within the observation channel and its value $\Phi_R$ in the reservoir at the entrance is assumed to be the same as in the micro-viscosimeter.  {This is justified by the fact that in the low viscosity suspending medium,  and at least until volume fractions of $\Phi_T\sim 25\%$, the relationship between $\Phi_R$ and $\Phi_T$ in the viscosimeter is close to that proposed by Pries et al. in a cylindrical tube of diameter $h$ (Fig. \ref{fig:PhiT}) --- which is a result similar to that on viscosities, shown in Fig. \ref{fig:physio} and discussed later on. As the observation channel, of aspect ratio 3.25, can be considered as an intermediate situation beween the Hele-Shaw cell and the cylinder, we can estimate that the relationship between both volume fractions should be similar. As the  logarithm of the intensity profiles  in the channel yields only  a profile that is proportional to the local volume fraction profiles (Beer-Lambert law), the values  $\Phi_T$ estimated from this relationship are used to rescale these  profiles. Noteworthy, the relationship between reservoir and tube volume fractions does not seem to depend much on the viscosity of the external medium while RBC dynamics does \cite{minetti19}, a fact for which we have no clear explanation.}

 \section{Results and discussion}
 
 \subsection{Viscosity}
 
 In discussions on the rheology of suspensions, a convenient  and widely used parameter is the relative viscosity, that evaluates the contribution of the particles to the dissipative mechanisms. Recalling that $\eta_1(\Phi_T)$ is the viscosity of the suspension with tube volume fraction $\Phi_T$, the relative viscosity is defined as $\hat{\eta}(\Phi_T)=\eta(\Phi_T)/\eta'_0(\Phi_T)$ where $\eta'_0(\Phi_T)$ is the viscosity of the suspending medium of the suspension. It differs from the initial viscosity $ \eta_0$ of the reference fluid in which the cells were suspended since there is always a residual volume of washing fluid (of viscosity $\eta_w\sim$ 1 mPa.s) in the centrifuged red blood cells. It has been estimated to be around 3\% for similar centrifugation speeds by Chaplin \textit{et al} \cite{chaplin52}. Assuming linear contribution of this residual fluid, the corrected viscosity for a suspension where a volume $\Phi_R$ has been mixed with a volume $1-\Phi_R$ of suspending fluid  is then $\eta'_0=((1-\Phi_R) \eta_0+0.03\Phi_R \eta_w)/(1-0.97\Phi_R)$. This leads to corrections in viscosity of up to 3\% for the more viscous fluid. The impact of the residual fluid after centrifugation on the final value of the volume fraction is, in contrast, much below the precision of haematocrit measurements.
 
 In principle, the microfluidic rheometer directly yields the ratio between the unknown viscosity of the suspension and that of the reference fluid. However, the finite thickness of the interface between the two fluids (the intensity profile necessarily varies on a scale of a few cell radii, {i.e. around 20 microns}) makes this measurement sensitive to the choice of a threshold level to determine the position of this interface. {This means that the absolute values of the relative viscosity that are found here are subjected to an offset of a few \% depending on the choice of the threshold for the position of the interface between the two fluids to be compared. Importantly, this offset does not vary much with $w_1/w_0$, meaning that this choice has no impact on the discussion about relative variations of viscosities, which is the goal of this paper.}

\begin{figure}
	\centering 		
\includegraphics[width=0.5\textwidth]{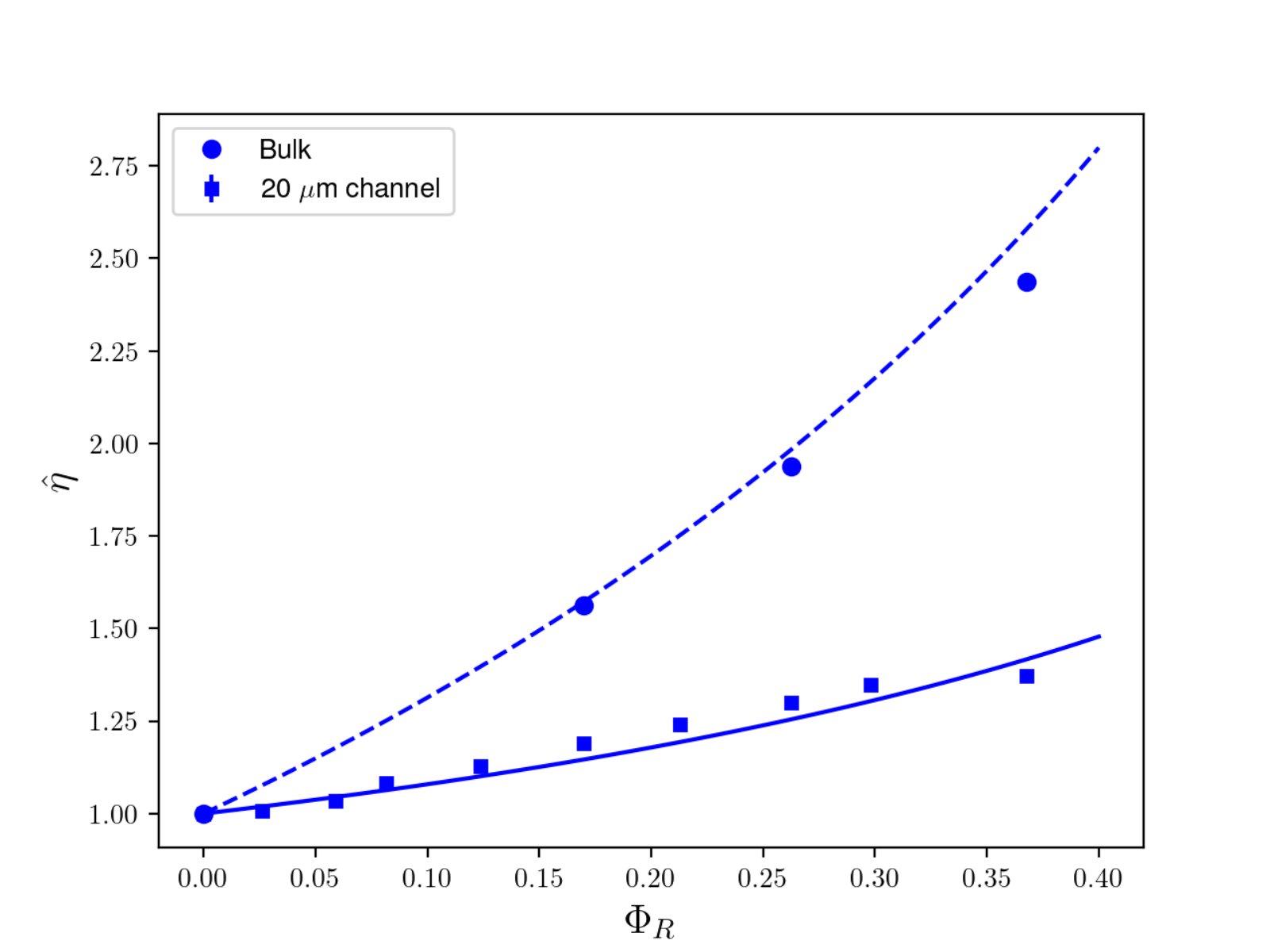}
	\caption{Relative viscosity $\hat{\eta}$ for $\eta_0=1.6$ mPa.s, as a function of reservoir volume fraction $\Phi_R$ in two geometries : 20 $\mu$m channel and cone-plate rheometer measuring the bulk viscosity. Lines correspond to the blood viscosity model given by Pries {\it et al.} \cite{pries92}, for cylindrical channels of 20 $\mu$m and infinitely large diameters, respectively, with no adjustable parameter.\label{fig:physio}}
\end{figure}

For a suspending fluid of viscosity similar to plasma in blood, a marked difference between confined and bulk rheology can be seen (Fig. \ref{fig:physio}). The decrease of viscosity with decreasing tube diameter (until the latter approaches RBC diameter, that is about {7.9 $\pm$ 0.7 $\mu$m \cite{linderkamp82}}) is known as the F\r{a}hr\ae{}us-Lindquist effect \cite{Lindqvist31} and has been thoroughly studied in the literature. The decrease is essentially explained by the existence of a cell depleted layer near the walls, acting as a lubrication film, and whose relative contribution to the overall dissipation becomes more and more important as the tube diameter decreases. An ad-hoc model accounting for the dependency of the viscosity with the cell volume fraction and the tube diameter has been proposed by Pries \textit{et al.}\cite{pries92}, which is now recognized as a correct description of blood apparent viscosity in tube flow. While our geometry is confined only in one dimension, we show in Fig. \ref{fig:physio} that this empirical law describes well our experimental data, indicating that a 2D structuring has a similar effect on rheology as in a 3D axi-symmetric situation.

\begin{figure}
	\centering 		
\includegraphics[width=0.5\textwidth]{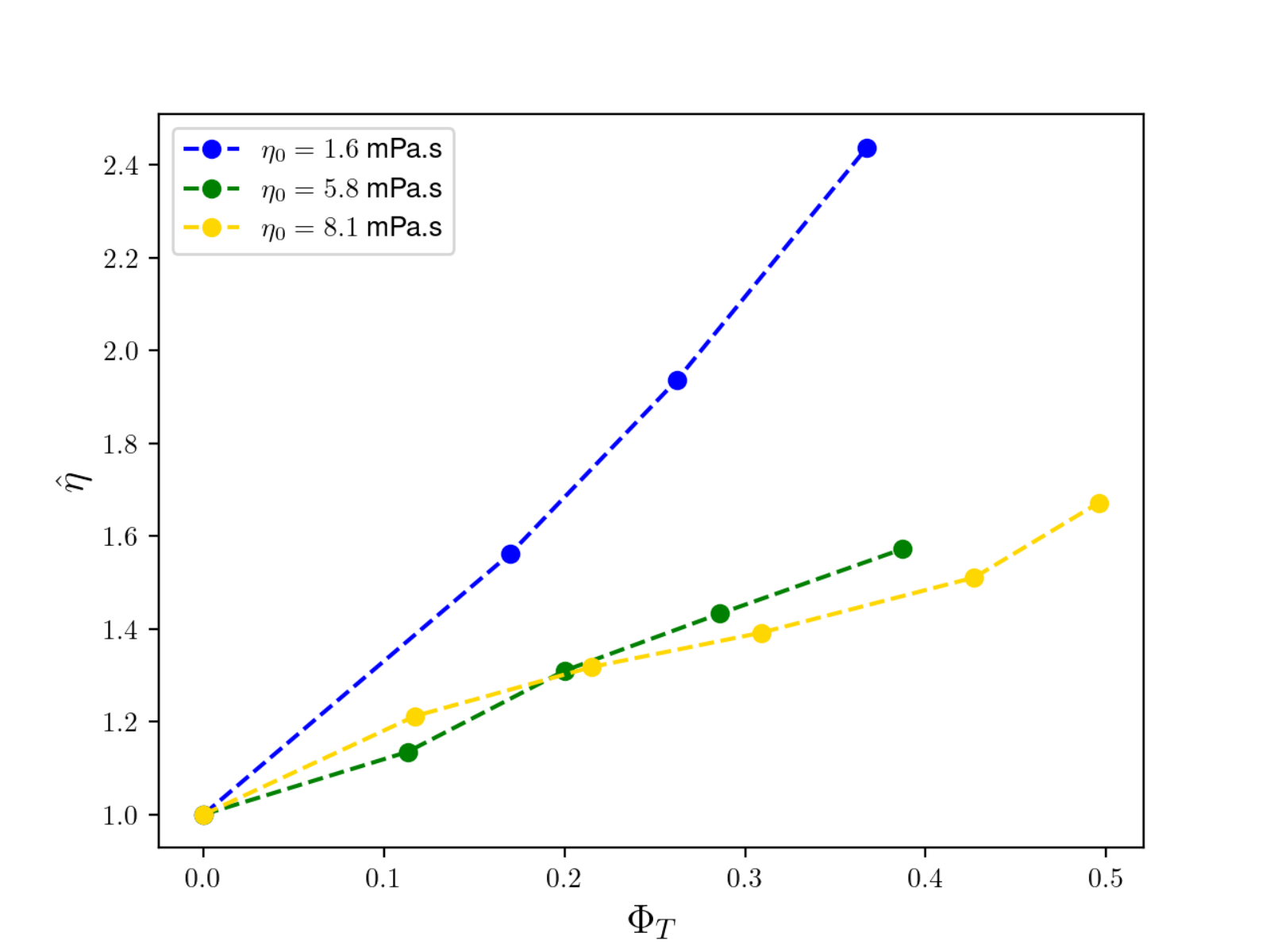}
	\caption{ Relative bulk viscosity $\hat{\eta}$  as a function of tube volume fraction $\Phi_T$ (here equal to reservoir volume fraction $\Phi_R$), for three different initial viscosities of the suspending fluid.\label{fig:CP}}
\end{figure}

If we now consider the three considered suspending media, Fig. \ref{fig:CP} shows that the contribution of cells to bulk rheology may depend on the viscosity of the suspending fluid; more precisely, the roughly linear behaviour $\hat{\eta}=1+[\hat{\eta}] \Phi_T$ of the relative viscosity  with  volume fraction is characterized by an intrinsic viscosity $[\hat{\eta}]$ which is about 3 times smaller when $\eta_0$ goes from 1.6 to 5.8 mPa.s. By contrast, this coefficient is almost unchanged when further increasing $\eta_0$ to 8.1 mPa.s. This is in overall agreement with the exhaustive study of Vitkova \textit{et al.}\cite{vitkova08} where the authors discuss the link between this intrinsic viscosity and the dynamics of red blood cells, which itself strongly depends on the viscosity contrast between their cytosol and the suspending medium\cite{fischer13,minetti19}. {They showed that when this viscosity contrast is varied, the intrinsic viscosity exhibits a minimum when the viscosity of the suspending medium is close to the viscosity of the cytosol, corresponding to a transition between flipping and tank-treading dynamics of the cells. In our experiments, the last two cases where $\eta_0=5.8$ and $8.1$ mPa.s are close to this minimum where RBCs are aligned with the flow, while for  $\eta_0=1.6$ mPa.s they undergo a more dissipative tumbling dynamics.}

  \begin{figure}
	\centering 		
\includegraphics[width=0.5\textwidth]{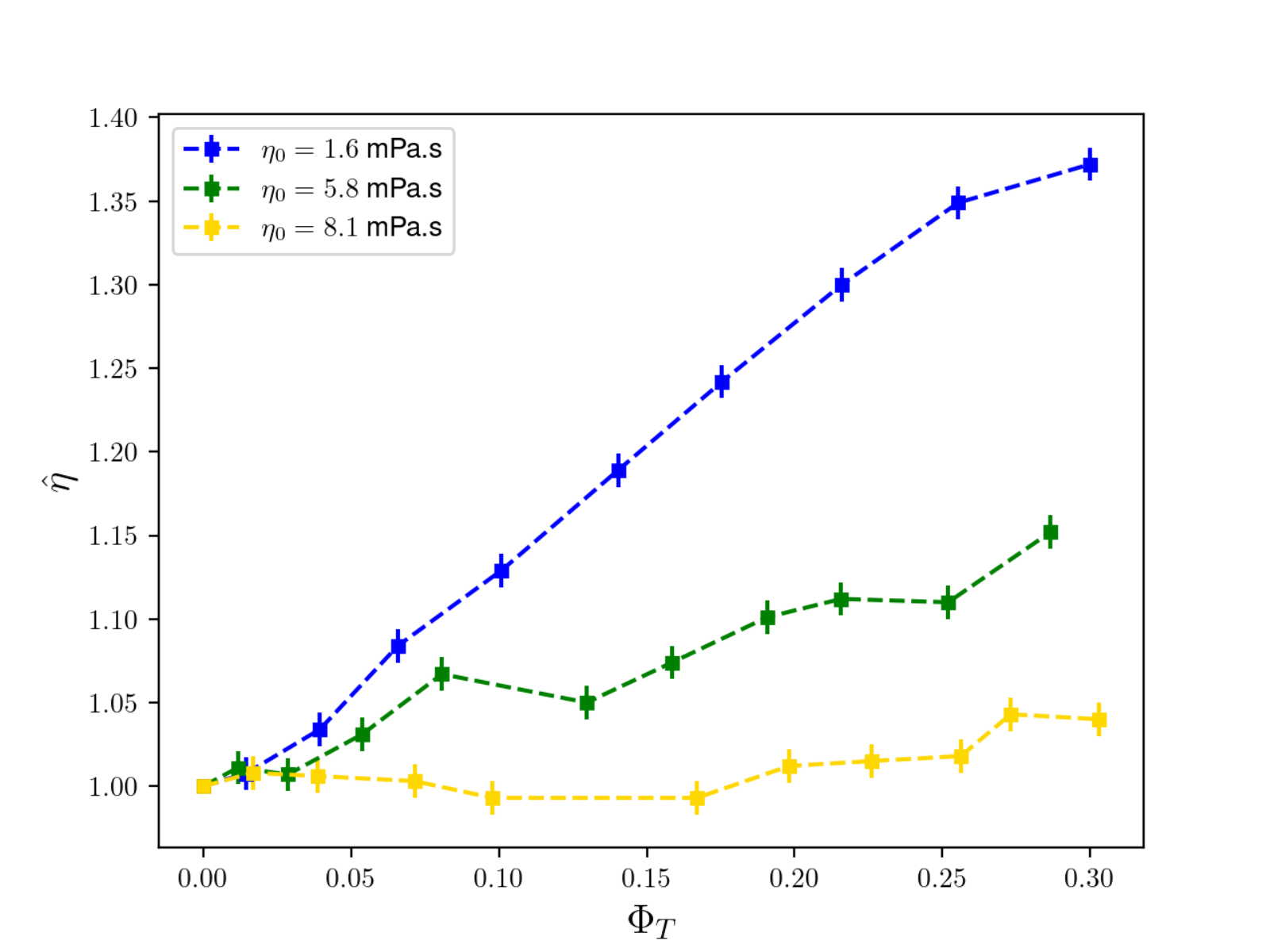}
	\caption{Relative viscosity $\hat{\eta}$ in the  20 $\mu$m channel  as a function of tube volume fraction $\Phi_T$, for three different initial viscosities of the suspending fluid.\label{fig:confined}}
\end{figure}

In the narrow channel, the evolution of viscosity with tube volume fraction is more complex (see Fig. \ref{fig:confined}). In the  $\eta_0=1.6$ mPa.s suspending medium a marked, steady increase of viscosity with volume fraction is observed. By contrast, in the $\eta=8.1$ mPa.s suspending medium, there is no significant variation of the viscosity over a large range of volume fraction $\Phi_T$, also in marked contrast to the scenario in bulk: remarkably, adding more particles to the suspension up to a volume fraction of 30 \% does not increase viscous dissipation and the effective viscosity by more than a few \%. In the $\eta_0=5.6$ mPa.s fluid, an intermediate scenario arises, with successions of increases with slopes similar to that in the less viscous fluid, followed by plateauing and even slight decreases within experimental errors, in particular around $\Phi_T=10 \%$.

These features are all the more significant as they are highlighted through $\hat{\eta}$, which compares two viscosities  measured with the same method.  The existence of a local minimum of viscosity around $\Phi_T=0.12$ is in line with the unpublished results of a recent PhD thesis \cite{hogan19}, where 2D simulations of cells having the same cytosol viscosity as that of the suspending fluid were run. In channels of width respectively 9, 10 and 12 $\mu$m, a local minimum in the relative viscosity is found for volume fractions around 0.3, 0.25  and 0.2, respectively. No minimum was observed for larger channels, but only a few volume fractions were considered, leaving the possibility for the minimum to be missed.

For a better understanding of this non-monotonic behaviour, we have observed and compared the structure of the suspension in the cases  $\eta_0=1.6$ mPa.s and 5.8 mPa.s.

\subsection{Structure}

  \begin{figure}
	\centering 		
\includegraphics[width=0.5\textwidth]{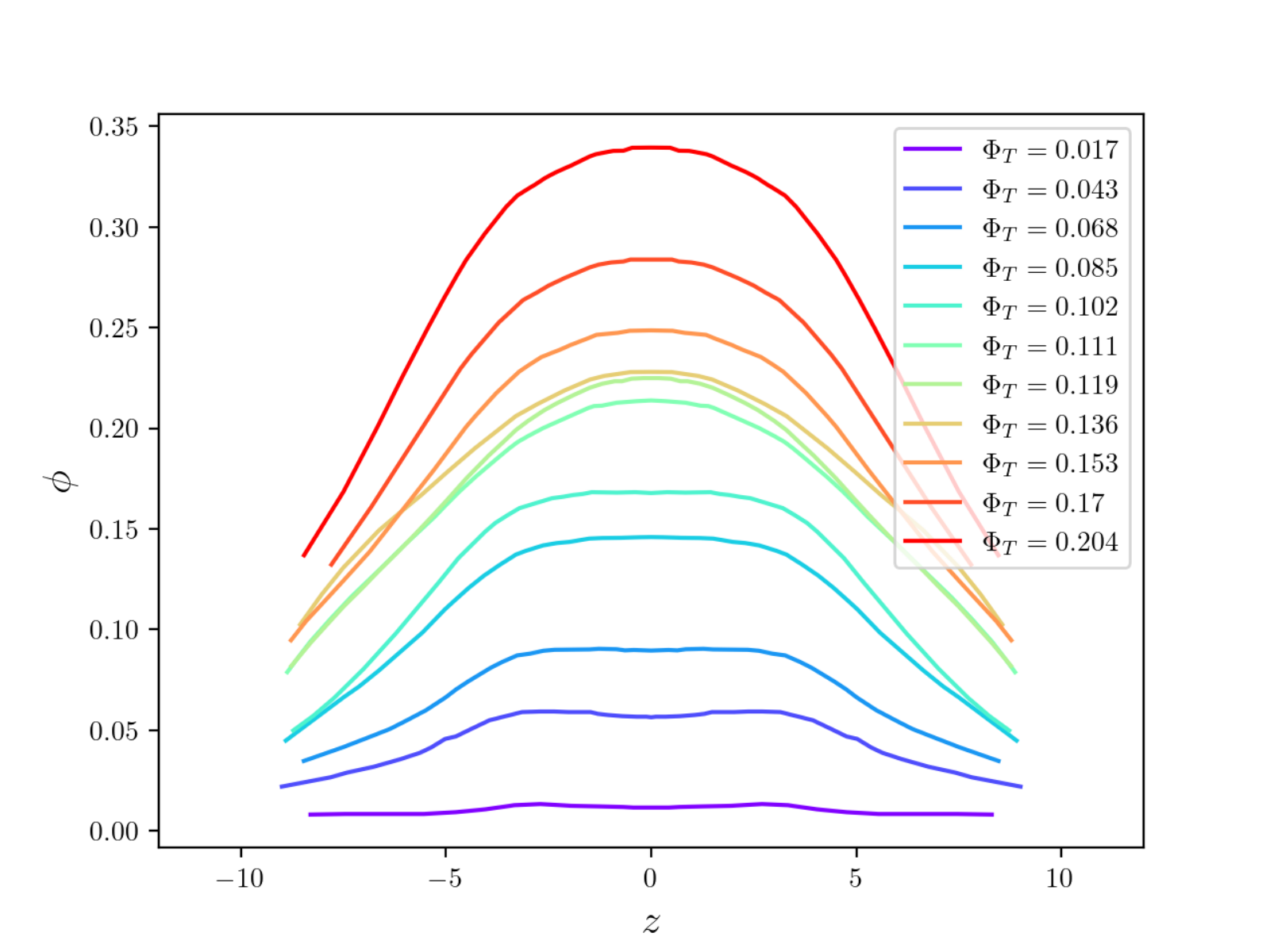}
	\caption{Concentration profiles in the $\eta_0=1.6$ mPa.s suspending fluid for different tube volume fractions $\Phi_T$.\label{fig:profil1}}
\end{figure}

  \begin{figure}
	\centering 		
\includegraphics[width=0.5\textwidth]{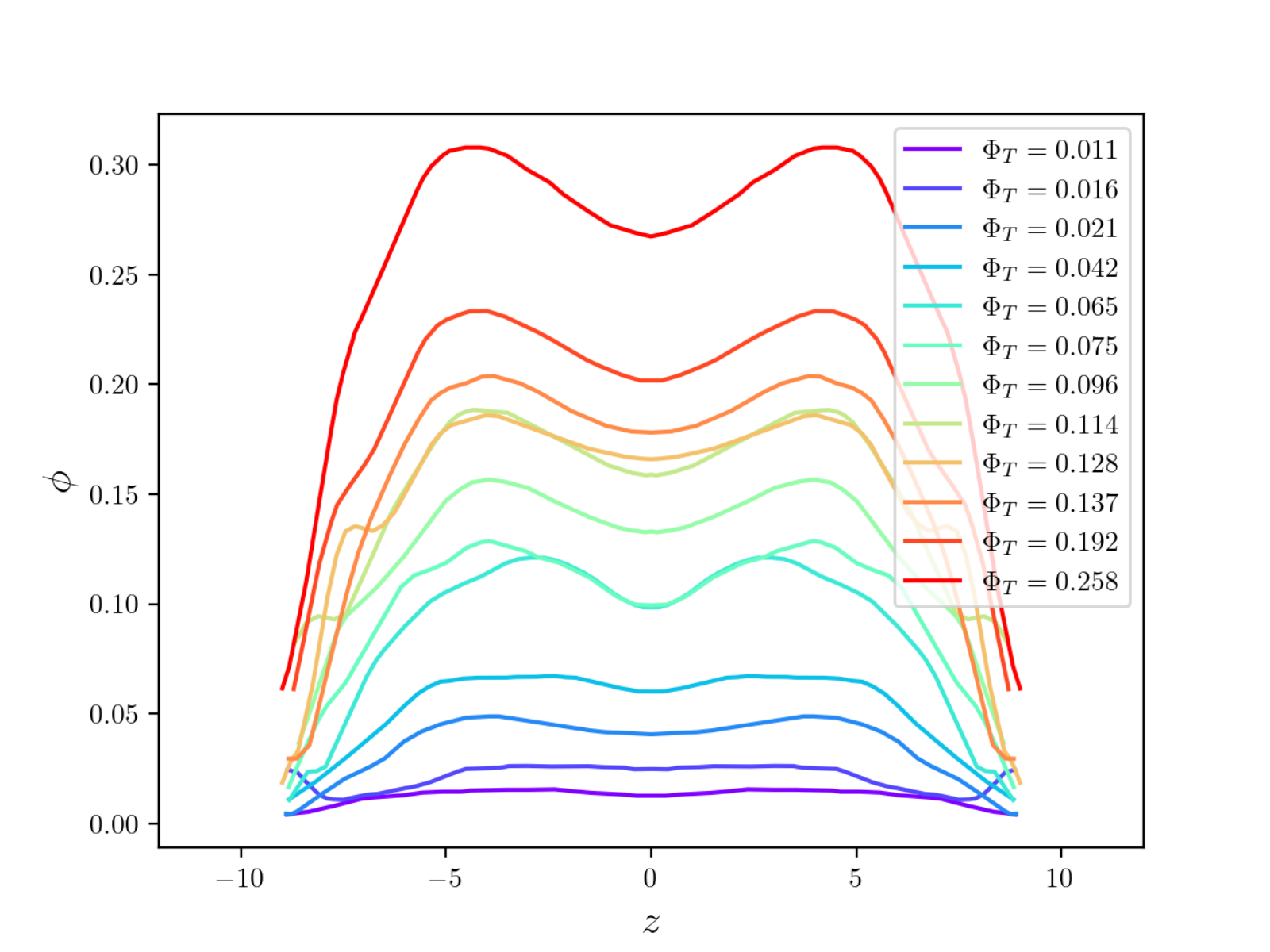}
	\caption{Concentration profiles in the $\eta_0=5.8$ mPa.s suspending fluid for different tube volume fractions $\Phi_T$.\label{fig:profil2}}
\end{figure}

In the low viscosity case, while a slight double-peak structure is visible at very low volume fractions, as in other works\cite{tsvirkun17}, a single peak, blunt profile quickly emerges as the volume fraction goes beyond 7\% (Fig. \ref{fig:profil1}). By contrast, in the  $\eta_=5.8$ mPa.s suspending medium, a marked double peak structure is present for almost all volume fractions  (Fig. \ref{fig:profil2}). Note that in all cases, the concentration between the two peaks is never 0, as can be sometimes seen in numerical simulations \cite{thiebaud14,shen17,feng21}. This can be related to limits of these numerical simulations, namely (i) they are 2D, a situation in which strong topological constraints impose that layers of particles, when they form, completely exclude each other while there are additional degrees of freedom in 3D simulations or experiments, in which particles do not need to be aligned in the same $(x,z)$ plane and (ii) most simulation works consider that all cells are identical (regarding their geometry and mechanical properties), while it is now well known that the dispersion in these properties (due in particular to ageing) leads to a dispersion of behaviour under flow \cite{fischer13,minetti19}, including the intensity of their migration velocity far from the channel walls \cite{losserand19}. Obtaining less pronounced structures in experiments was therefore expected.

\begin{figure}
	\centering 		
\includegraphics[width=0.5\textwidth]{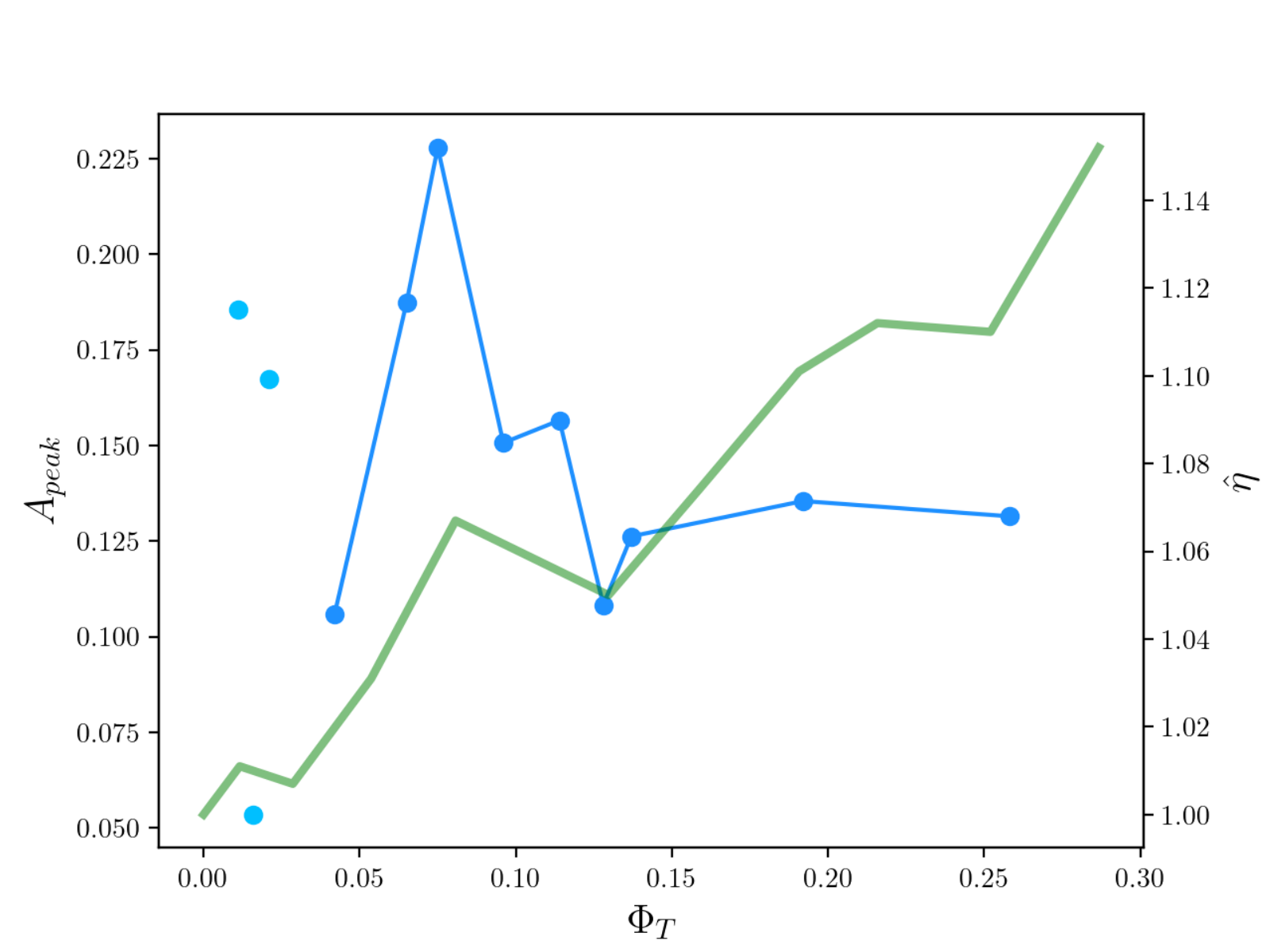}
	\caption{Blue dots, left axis: relative amplitude of the two peaks of the concentration profiles in the $\eta_0=5.8$ mPa.s suspending fluid. Faded points correspond to imprecise results at very low volume fractions for which the variations of concentration lead to fluctuations of only a few gray-levels in the images. The relative viscosity of the same suspension, similar to that of Fig. \ref{fig:confined}, is also plotted in full green line (right axis).\label{fig:peak}}
\end{figure}

We can characterize the importance of the double-peak structure by analysing the difference between the height of the peak and the concentration in the middle of the channel ($z=0$), rescaled by the peak value.
This relative amplitude $A_{peak}$ of the peaks is plotted in Fig. \ref{fig:peak} as a function of tube volume fraction.

We observe that in the 0.02-0.075 range of $\Phi_T$, the net increase in viscosity coincides with the formation of a marked double-peak structure (Fig. \ref{fig:peak}) and a significant decrease of the thickness of the cell-free layer near {top and bottom} walls seen in Fig. \ref{fig:profil2}, {as roughly evaluated by considering the position of the mid-height point}. Overall, the formation of this double-peak structure widens the distribution in comparison to what is seen in Fig. \ref{fig:profil1}. While the relative dissipation at the particle level should decrease when going from $\eta_0=1.6$~mPa.s to $\eta_0=5.8$~mPa.s due to different dynamics, as seen in bulk rheology (Fig. \ref{fig:CP}), the structuring effects (double peak and decreasing cell-free layer) actually lead to relative viscosities that are comparabe for  $\eta_0=5.8$mPa.s in this range of $\Phi_T$ as seen in Fig. \ref{fig:confined}.
For $0.075<\Phi_T<0.15$, the relative attenuation of the double peak structure is accompanied by a plateau or apparent small decrease in the measured viscosity. One can indeed suspect that adding more cells in the central part, where the velocity gradients are probably small, has only a small effect on viscous dissipation. In the $0.15<\Phi_T<0.25$ range, while it is more difficult to reach a conclusion as the trends are less marked, as also the case in simulations in  simple shear flow \cite{thiebaud14}, one can still notice, again, that in Fig. \ref{fig:profil2} a significant increase of the width of the distribution or decrease of the cell free layer takes place between $\Phi_T=0.137$ and $\Phi_T=0.192$, corresponding to a sharp increase of the viscosity, while  between $\Phi_T=0.192$ and $\Phi_T=0.258$ only the height of the profile varies, corresponding to a slower increase of the viscosity. One can suspect that in this range, the velocity profile is essentially a blunt one in the central part, with little contribution to the overall dissipation.

While it would be interesting to extend and confirm these results, especially by investigating the structure-rheology relationship at different values of the channel thickness $h$, the trends observed here are in line with recent simulations in 2D Poiseuille flow\cite{feng21} and show marked changes in the evolution of the effective viscosity with particle volume fraction. These are associated to structural transitions from one layer to two layers, the relative amplitude of these two peaks and the width of the overall distribution.

 \section{Conclusions}
 
We performed an experimental study of the rheology and structure of a suspension of {3D} deformable particles - red blood cells - flowing in a quasi-2D Poiseuille flow in a narrow channel. Combining a precise microfluidic viscosimetry technique and measurements of the particle concentration profiles in the thickness of the channel, we highlighted the evolution of the structure and viscosity of the suspension when varying a) the volume concentration of particles  and b)  their dynamics (and subsequent ability to organize into layers) by varying the viscosity of the suspending fluid.

While a general feature is that the effective viscosity of suspensions is significantly lower in confined channel flow than in bulk shear flow due to the well-know existence of a particle-free lubricating layer near walls, noticeable qualitative and quantitative differences are revealed when the viscosity of the suspending medium is increased.

At low suspending medium viscosity, a situation {close to physiological conditions} in which the particles are effectively less deformed by the flow and exhibit a tumbling dynamics in an unbounded shear flow, a simple structure takes place. Particle migration towards the centerline and shear-induced diffusion due to particle-particle interactions lead to a bell-shaped concentration profile with a maximum at the center and cell-free layers near walls, this profile being almost self-similar when the average concentration is increased. This is associated to a nearly steady, monotonic increase of the relative viscosity when the particle volume fraction is increased, with an average slope of about 1.

When the viscosity of the suspending fluid is increased, the situation becomes more complex and nontrivial. At an intermediate value of this viscosity (and thus of particle deformability), strong structuring effects take place in which the system shows a transition between a 1-peak concentration profile (as in the previous case) to a 2-peak structure in which hydrodynamic repulsion between particles leads to a local minimum of the concentration along the centerline. This qualitative change in the structure of the suspension as well as irregular variations of the thickness of the particle-free layer near walls lead to strong fluctuations of the evolution of effective viscosity with concentration, with a succession of ranges of rapid growth and plateaus (that may even suggest small decreases of the viscosity). At even higher suspending fluid viscosity, the impact of particle dynamics is even more drastic and a variation of only 3\% of the effective viscosity is measured for particle volume fractions up to 30\%.

While a few predictions on the relationship between strong structuring effects and rheology have been made in confined simple shear flow \cite{thiebaud14,shen17} or more recently in Poiseuille flow \cite{feng21}, these numerical simulations were made in pure 2D situations where topological constraints intuitively lead to stronger structuring effects due to the smaller number of degrees of freedom. To our knowledge, our experimental study provides an important confirmation that such effects exist and are measurable {with 3D particles confined in only one direction}  and qualitatively confirms the predictions made on capsules in 2D-Poiseuille flow~\cite{feng21}. This study, which certainly deserves to be extended to a wider range of parameters and possibly to other physical systems of deformable particles, provides interesting perspectives on the structure-rheology relationship of suspensions in channel flows.

\begin{acknowledgments}
V. A. acknowledges a PhD fellowship from "Ecole Doctorale de Physique de Grenoble". T. P. and G. C. acknowledge funding from CNES on blood rheology and flows.The authors' team is part of LabEx Tec 21 (Investissements d'Avenir - grant agreement ANR-11-LABX-0030).
\end{acknowledgments}

\section*{Data availability}

The data that support the findings of this study are available from the corresponding author upon reasonable request.


\end{document}